\documentclass[sigconf]{acmart}

\AtBeginDocument{%
  \providecommand\BibTeX{{%
    \normalfont B\kern-0.5em{\scshape i\kern-0.25em b}\kern-0.8em\TeX}}}

\setcopyright{acmcopyright}
\copyrightyear{2022}
\acmYear{2022}

\acmConference{CIKM '22}
\acmISBN{978-1-4503-XXXX-X/18/06}

\usepackage{paralist}
\usepackage{algorithm}
\usepackage{algorithmic}

\usepackage{amsmath}
\usepackage{multirow}

\usepackage{subfigure}
\usepackage{graphicx}
\usepackage{epsfig}
\usepackage{epstopdf}

\usepackage{bm}
\usepackage{booktabs}
\usepackage{verbatim}
\usepackage{array}

\begin{document}

\title{Time Lag Aware Sequential Recommendation}
%\title{Multi-Sparse-Domain Collaborative Recommendation via Comprehensive Aspect Preference Learning}

\author{Lihua Chen} 
\affiliation{%
\institution{School of Computer Science}
  \country{Sichuan University, China}
}
\email{clhua@outlook.com}

\author{Ning Yang}
\authornote{Corresponding Author.}
\affiliation{%
\institution{School of Computer Science}
  \country{Sichuan University, China}
}
\email{yangning@scu.edu.cn}

\author{Philip S. Yu}
\affiliation{%
\institution{Department of Computer Science}
  \country{University of Illinois at Chicago, USA}
}
\email{psyu@uic.edu}



\begin{abstract}
Although a variety of methods have been proposed for sequential recommendation, it is still far from being well solved partly due to two challenges. First, the existing methods often lack the simultaneous consideration of the global stability and local fluctuation of user preference, which might degrade the learning of a user's current preference. Second, the existing methods often use a scalar based weighting schema to fuse the long-term and short-term preferences, which is too coarse to learn an expressive embedding of current preference. To address the two challenges, we propose a novel model called Time Lag aware Sequential Recommendation (TLSRec), which integrates a hierarchical modeling of user preference and a time lag sensitive fine-grained fusion of the long-term and short-term preferences. TLSRec employs a hierarchical self-attention network to learn users' preference at both global and local time scales, and a neural time gate to adaptively regulate the contributions of the long-term and short-term preferences for the learning of a user's current preference at the aspect level and based on the lag between the current time and the time of the last behavior of a user. The extensive experiments conducted on real datasets verify the effectiveness of TLSRec. 
\end{abstract}






\maketitle

\section{Introduction}
In recent years, sequential recommendation, also known as session-based or sequence-aware recommendation, has been attracting increasing interest of researchers \cite{Quadrana2018,Fang2019}. Sequential recommender systems aim to capture the time-sensitive preference (or needs) of users by modeling the sequential dependency between their behaviors based on their historical interaction data (e.g., click, purchase, and check-in) that are collected sequentially by online platforms such as e-commerce websites and location-based networks. The information about the sequential dependency and time-sensitive preference can be used for applications where items need to be recommended based on a user's previous interactions. For example, a sequential recommender system can timely recommend AirPod to a user after she/he purchases an iPhone.

\subsection{Related Work}

A variety of methods have been proposed for sequential recommendation. Early works are often based on Markov chain which assumes each interaction highly depends on its previous ones \cite{Quadrana2018,Rendle2010,He2016,xie2021adversarial,xia2021self}. Recently, inspired by the impressive success of deep learning techniques in the fields of natural language processing and computer vision, lots of deep learning based models have also been proposed for sequential recommendation and achieved the state-of-the-art performance \cite{Zhang2019,Fang2019}. Early deep learning based methods utilize recurrent neural networks (RNN) to characterize the dynamics of interaction sequences \cite{Hidasi2016,Hidasi2018,Cui2020}, which however suffers from inability to capture the long-term dependency between interactions, i.e., one interaction likely depends not only on the recent interactions but also on early ones. To overcome this drawback, another line of deep learning based methods employs attention mechanism \cite{Vaswani2017,Li2017,Ting2018,Ying2018,Sun2019,ChenCIKM2019,Li2020,Tanjim2020,Ren2020,Wu2020,Li2021} and graph neural network (GNN) \cite{hsu2021retagnn,chang2021sequential,wang2020next,xia2021self} to model sequential dependency relationships between interactions and identify relevant items.

\subsection{Challenges}
Notwithstanding the improvements on sequential recommendation, it is still far from being well solved partly due to the following two challenges.

\begin{compactitem}

\item \textbf{Unification of Stability and Fluctuation of Preferences} In sequential recommender systems, user behaviors are often organized into sessions or transactions and basically driven by a mix of two factors, long-term preference and short-term preference. The long-term preference reflects a user's general interest which usually changes slowly and keeps relative stable across sessions, while the short-term preference represents a user's taste in a session which might deviate from her/his long-term preference \cite{Jannach2015,Song2016,Feng2019}. For example, a user usually prefers to "classic music", but probably in some days she/he is particularly fond of "rock and roll" because of the influence of her/his friends. However, the existing methods for sequential recommendation often treat the user preference as a flat distribution over sessions, without distinguishing its global stability and local fluctuation, which might degrade the learning of user preference. We need a method which can capture the stability of long-term preference at global time scale, as well as the fluctuation of short-term preference at local time scale. 

\item \textbf{Fine-grained Fusion of Long-term and Short-term Preferences} To make effective sequential recommendations, it is extremely important to simultaneously capture users' long-term preferences across different sessions and their short-term preferences in recent sessions, so that the current preference of users can be learned. Recently, some sequential recommendation models have been proposed for fusing the embeddings of long-term preference and short-term preference with static weights as hyper-parameters \cite{Song2016,An2019} or using dynamic attentional coefficients produced by an attention network \cite{Feng2019}. However, no matter whether the static weights or the dynamic attentional coefficients they use, in the existing models a preference embedding vector is weighted by a scalar, which implicitly assumes that different dimensions in the same preference embedding have the same importance. We argue that such scalar based weighting scheme is too coarse for learning an expressive fused preference embedding, as in real world, a user's behaviors might depend more on some aspects than on other aspects of preference. For example, a user might buy a science fiction book because the genre aspect of her/his long-term preference to movie is science fiction and she/he currently likes reading. In this example, the genre aspect  should be weighted more than other aspects of the long-term preference. Therefore, we need a more fine-grained fusing mechanism that can adaptively capture the different contributions of different aspects of long-term preference and short-term preference for the fusion of them.

\end{compactitem}

\subsection{Contributions}

To address the above challenges, we propose a novel model called Time Lag aware Sequential Recommendations (TLSRec), which integrates a hierarchical modeling of user preference and a time lag sensitive fine-grained fusion of the long-term and short-term preferences. At first, in contrast with the traditional sequential recommendation methods that capture the long-term preference directly from the flat sequence of interactions without considering the preference fluctuation between local sessions, TLSRec can simultaneously model the global stability and local fluctuation of a user's preference with a hierarchical self-attention network consisting of a short-erm preference learning layer and a long-term preference learning layer. Such unified modeling offers TLSRec the ability to understand a user's preference at both local and global time scales. Particularly, in order to capture the preference fluctuation between local sessions, TLSRec learns a session embedding for each session to encode a user's preference local to a session, with a self-attention module at the short-term preference learning layer. At the same time, in order to capture the intrinsic stable preference of a user, TLSRec will pool the session embeddings into a long-term preference embedding with a multi-head self-attention module at the long-term preference learning layer. Due to the different self-attention heads, not only the long-term dependency between sessions but also the interactions between dimensions of session embeddings can be perceived by the long-term preference embedding to enhance its ability to capture the stable intrinsic preference.

%Due to the self-attention module, the session embeddings can capture the dependency between the items in a session and identify their different contributions to a user's short-term preference. Particularly, TLSRec uses the embedding of the last session to represent the current short-term preference of a user. In order to capture the intrinsic stable preference of a user, TLSRec will pool the session embeddings into a long-term preference embedding at the long-term preference learning layer. In contrast to the existing works that are often based on single-head self-attention, TLSRec uses a multi-head self-attention module for the learning of the long-term preference embedding. Due to the different self-attention heads, not only the long-term dependency between sessions but also the interactions between dimensions of session embeddings can be perceived by the long-term preference embedding to enhance its ability to capture the stable intrinsic preference.

To overcome the challenge of fine-grained fusion of the long-term preference and short-term preference for sequential recommendations, inspired by the idea of the gates in long short-term memory (LSTM) \cite{hochreiter1997long,Chen2019}, we propose a neural time gate for TLSRec to learn a fused preference embedding aware of time lag. Compared with traditional methods which weigh the long-term preference and short-term preference with manually defined scalars as vector-wise weights, the proposed neural time gate has two advantages. First, it offers TLSRec the ability to adaptively regulate the contributions of the long-term preference and the short-term preference based on the time lag. The idea here is that which preference accounts more for a user's next behavior heuristically depends on the \textbf{time lag}, i.e., how long has it been since her/his last behavior. As we will see in later experiments, the neural time gate will learn to act in accordance with the intuition that the longer (shorter) the time lag, the more the impact of a user's long-term (short-term) preference on her/his next behavior. Second, in contrast with the existing works, the neural time gate offers a fusion of the long-term preference and short-term preference at a finer granularity level. Unlike the existing works, the neural time gate will generate a gating vector instead of a scalar, whose dimensions serve as dimension-wise weights to differentially weigh the corresponding dimensions of the long-term embedding and short-term embedding. Due to the gate based fusion of the long-term preference and short-term preference, TLSRec can learn a more representative and comprehensive hybrid embedding for current preference. Finally, the contributions of this paper can be summarized as follows:
\begin{compactitem}
    \item We propose a novel model called Time Lag aware Sequential Recommendations (TLSRec), which can capture the stability and fluctuation of user preference, and learn a fused preference embedding with a fined-grained fusion of the long-term preference and short-term preference.
    
    \item We propose a hierarchical self-attention network to unite the learning of the long-term preference and short-term preference, which leads to a better comprehension of the stability of user preference at global time scale as well as the fluctuation of user preference at local time scale.
    
    \item We propose a neural time gate to offer a gate based fine-grained fusion of the long-term preference and short-term preference at the dimension level, by which a more representative and more comprehensive fused preference embedding can be learned.
        
     \item We extensively evaluate TLSRec on real-world datasets. The experimental results demonstrate the general improvements of TLSRec over the baselines, as well as the effectiveness of the proposed hierarchical self-attention network and neural time gate.
\end{compactitem}

%The rest of this paper is organized as follows. We give the preliminaries in Section 2. The details of TLSRec are described in Section 3. In Section 4 we analyze the experimental results. Finally, we briefly review the related work in Section 5 and conclude in Section 6.

\section{Preliminaries}
%In this section, we first define some basic notations, then formulate the target problem of this paper.

%\begin{table}[t]
%\centering
%  \caption{Notations}
%  \label{tab:notation}
%  \begin{tabular}{ll}
%    \hline
%    Notation &Description\\
%    \midrule
%    $\mathcal{U}$, $\mathcal{V}$ & user and item set\\
%    $N$, $M$ & the number of users and items \\ 
%    $u$, $v$ & user and item\\
%    $\mathcal{S}^u$ & session sequence of user $u$ \\
%    $\mathcal{S}^u_i$ & the $i$th session of user $u$ \\
%    $T$ & the number of sessions of user $u$\\
%    $m$ & the number of items in a session \\
%    $v^u_{i,j}$ & the $j$th item user $u$ interacts with in $i$th session\\
%    $t(v)$ & the time when interaction with item $v$ happens\\
%    $\boldsymbol{e}$ & item embedding\\
%    $\boldsymbol{f}$ & user embedding\\
%   $\boldsymbol{s}$ & session embedding\\
%    $\boldsymbol{p}$ & position embedding\\
%    $\boldsymbol{z}^u_{\text{long}}$ & the long-term preference embedding of user $u$\\
%    $\boldsymbol{z}^u_{\text{short}}$ & the short-term preference embedding of user $u$\\
%    $\boldsymbol{y}$, $\boldsymbol{g}$ & the time lag embedding and the time gate vector\\  
%    $r_{u,v}$ & the rating of user $u$ on item $v$\\
%    $[\boldsymbol{x}_i]_{i=1}^{T}$ & matrix composed of $T$ column vectors $\boldsymbol{x}_i$\\
%  \hline
%\end{tabular}
%\end{table}

%\subsection{Basic Definitions}
Let $\mathcal{U}$ be a set of $N$ users and $\mathcal{V}$ a set of $M$ items. The interactions of user $u \in \mathcal{U}$ sorted chronologically is organized as a sequence of $T$ sessions $\mathcal{S}^u = \langle \mathcal{S}^u_1, \mathcal{S}^u_2, \cdots, \mathcal{S}^u_{T} \rangle$. Each session $\mathcal{S}^u_i = \{ v^u_{i,1}, v^u_{i,2}, \cdots, v^u_{i, |\mathcal{S}^u_i|}\}$ is a subset of $\mathcal{V}$, where $v^u_{i,j} \in \mathcal{V}$ ($1 \le j \le |\mathcal{S}^u_i|$) is the $j$th item user $u$ interacts with in the session $\mathcal{S}^u_i$. Let $t(v)$ be the time when the interaction with item $v \in \mathcal{V}$ happens, and then for any $v^u_i \in \mathcal{S}^u_i$ and any $v^u_j \in \mathcal{S}^u_j$, $t(v^u_i) < t(v^u_j)$ if $i < j$. 
%The main notations are listed in Table \ref{tab:notation}.

%\subsection{Problem Formulation}
Given a user $u \in \mathcal{U}$ and her/his historical sessions $\mathcal{S}^u = \langle \mathcal{S}^u_1, \mathcal{S}^u_2,$ $ \cdots, \mathcal{S}^u_{T} \rangle$, we want to recommend $k$ items that $u$ will most probably interact with in the next session $\mathcal{S}^u_{T+1}$. This problem can be formulated as a ranking problem of all items for user $u$ based on the rating prediction $\hat{r}_{u, v}$ of user $u$ over item $v \in \mathcal{V}$. 

\begin{figure*}[!t]
    \centering
    \includegraphics[width=0.6\textwidth]{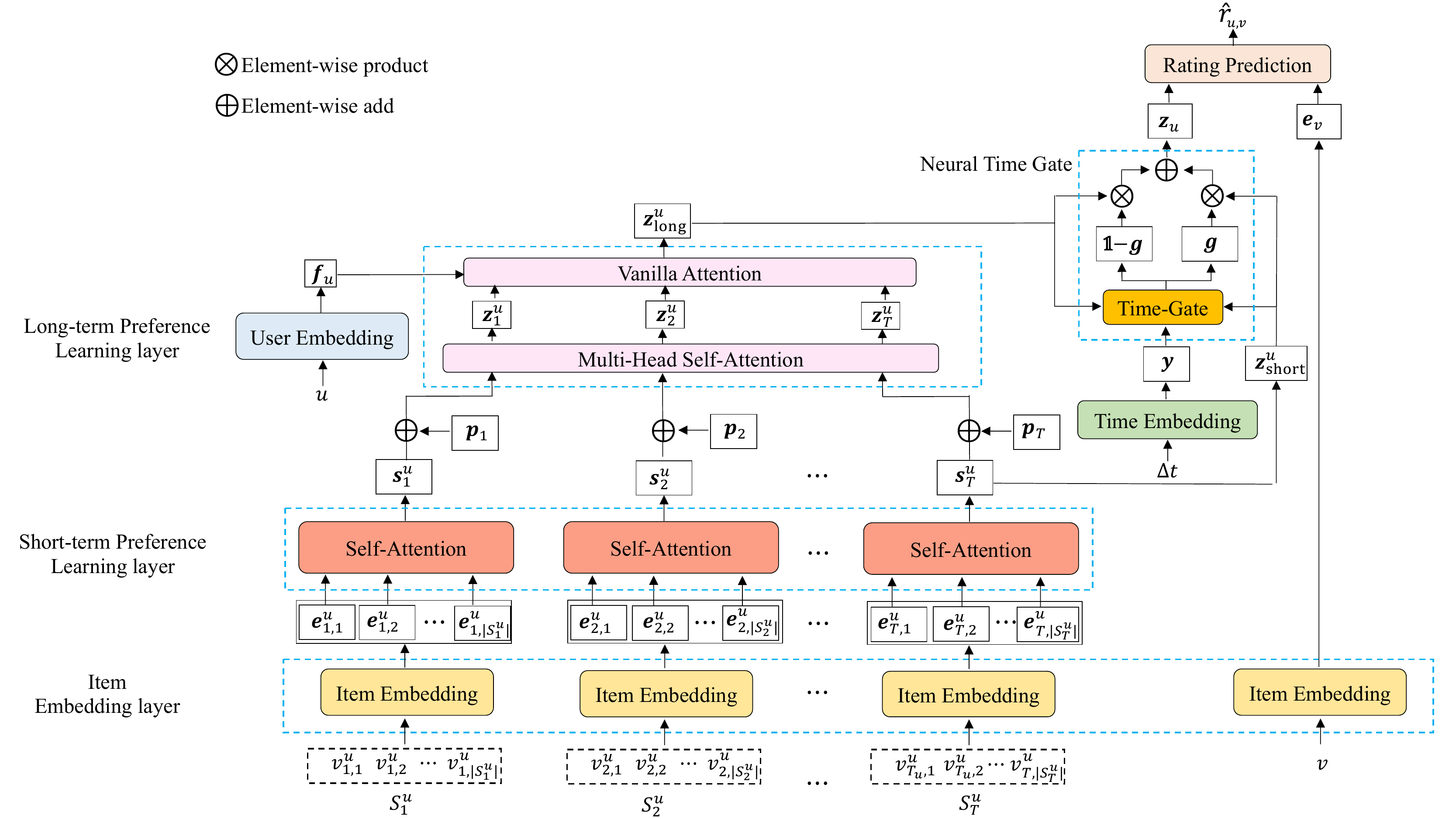}
    \caption{The architecture of TLSRec.}
    \label{fig:model}
\end{figure*}

\section{Proposed Model}
%In this section, we first give an overview of the proposed TLSRec model, then present its components in detail, and finally describe its learning.

\subsection{Overview of TLSRec}

The architecture of TLSRec is shown in Figure \ref{fig:model}. As we can see from Figure \ref{fig:model}, given as inputs the historical session sequence $\langle \mathcal{S}^u_1, \mathcal{S}^u_2, \cdots, \mathcal{S}^u_{T} \rangle$ of a specific user $u$, the lag $\Delta t$ between the time when the recommendation is made and the time of the last interaction of $u$, and a candidate item $v$,  TLSRec is supposed to produce the predicted rating $\widehat{r}_{u,v}$ of $u$ on $v$. 

First, TLSRec uses an $M$-dimensional one-hot vector to encode an item, and transforms each item $v^u_{i,j}$ in each session $\mathcal{S}^u_i$ to its corresponding item embedding $\boldsymbol{e}^u_{i,j} \in \mathbb{R}^{d}$ through the item embedding layer, where $d$ is the dimensionality of embeddings, $1 \le i \le T$, and $1 \le j \le |\mathcal{S}^u_i|$. 

Then at the short-term preference learning layer, the item embeddings in a session $\mathcal{S}^u_i$ will be aggregated into its corresponding session embedding $\boldsymbol{s}^u_i \in \mathbb{R}^{d}$ with a self-attention module shared across sessions. The session embedding $\boldsymbol{s}^u_i$ encodes user $u$'s short-term preference which is local to session $\mathcal{S}^u_i$, and the differences between them reflect the fluctuation of user preference among short time periods. At the same time, note that the current short-term preference embedding $\boldsymbol{z}^u_{\text{short}}$ is just the same as the last session embedding $\boldsymbol{s}^u_{T}$, as the current preference of a user is often revealed by the interactions in her/his most recent session \cite{Yu2016,Ying2018}.

The task of the long-term preference learning layer is to generate the long-term preference embedding $\boldsymbol{z}^u_{\text{long}} \in \mathbb{R}^{d}$ by fusing the session embeddings with a multi-head self-attention module. As we have mentioned before, the multiple self-attentional heads enable TLSRec to capture the interactions between dimensions of session embeddings, which leads to the more representative and comprehensive attentional session embeddings $\boldsymbol{z}^{u}_{i} \in \mathbb{R}^{d}$ at the global time scale. Meanwhile, as the same as the existing transformer-based models do \cite{Vaswani2017,Kang2018,Ren2020}, TLSRec will incorporate the session embeddings with a learnable position embedding $\boldsymbol{p}_i \in \mathbb{R}^{d}$ ($1 \le i \le T$) before feeding them into the multi-head self-attention module, so that the temporal dependency between sessions can be perceived by the long-term preference embedding. At last, the attentional session embeddings $\boldsymbol{z}^{u}_{i}$ are fused into the long-term preference embedding $\boldsymbol{z}^u_{\text{long}}$ via a vanilla attention module, by which the long-term preference can be aware of the different contributions of different sessions.

Once the long-term preference embedding $\boldsymbol{z}^u_{\text{long}}$ and the short-term preference embedding $\boldsymbol{z}^u_{\text{short}}$ are prepared, TLSRec will merge them through the neural time gate to generate the final preference embedding $\boldsymbol{z}_u$. For regulating the contributions of the long-term preference and the short-term preference, the neural time gate will generate an intermediate gate vector $\boldsymbol{g} \in \mathbb{R}^d$ based on the time embedding $\boldsymbol{y} \in \mathbb{R}^d$ of the time lag $\Delta t$ to adaptively weight the dimensions of $\boldsymbol{z}^u_{\text{long}}$ and $\boldsymbol{z}^u_{\text{short}}$. At last, TLSRec will make the rating prediction $\widehat{r}_{u,v}$ based on the inner product of the preference embedding $\boldsymbol{z}_u$ and the item embedding $\boldsymbol{e}_{v}$ of the candidate item $v$.

\subsection{Hierarchical Self-Attention Network}

\subsubsection{Item Embedding}
For any item $v \in \mathcal{V}$, we obtain its embedding $\boldsymbol{e}_{v} \in \mathbb{R}^d$ by a lookup over a learnable matrix $\boldsymbol{W}^{\text{I}} \in \mathbb{R}^{d \times M}$, i.e., $\boldsymbol{e}_{v} = \boldsymbol{W}^{\text{I}} \boldsymbol{v}$,
%\begin{equation}
%\boldsymbol{e}_{v} = \boldsymbol{W}^{\text{I}} \boldsymbol{v},
%\label{Eq_Item_Embedding}
%\end{equation}
where $\boldsymbol{v} \in \mathbb{R}^M$ is a one-hot vector representing the item $v$. For the items $ \{ v^u_{i,1}, v^u_{i,2}, \cdots, v^u_{i,m}\}$ of session $\mathcal{S}^u_i$ of a user $u$, we horizontally assemble their item embeddings into an item embedding matrix $\boldsymbol{E}_i = [\boldsymbol{e}^u_{i,j} ]_{j=1}^{m} \in \mathbb{R}^{d \times m}$, where $m = |\mathcal{S}_i|$ and the $j$th column $\boldsymbol{e}^u_{i,j} \in \mathbb{R}^d$ ($1 \le j \le m$) is the embedding of item $v^u_{i,j}$. 
%obtained by Equation (\ref{Eq_Item_Embedding}).

\subsubsection{Short-term Preference Learning}
Given a user $u \in \mathcal{U}$ and her/his historical session sequence $\mathcal{S}^u = \langle \mathcal{S}^u_1, \mathcal{S}^u_2, \cdots,$ $ \mathcal{S}^u_{T} \rangle$, the task of the short-term preference learning layer is to generate the session embeddings $\boldsymbol{s}^u_i$ representing $u$'s preference local to each session $\mathcal{S}^u_i$, $1 \le i \le T$, using a self-attention module. For this purpose, each item embedding $\boldsymbol{e}^u_{i,j} $ in session $\mathcal{S}^u_i$ will first be transformed to three vectors, a query vector $\boldsymbol{q}_{i,j} \in \mathbb{R}^d$, a key vector $\boldsymbol{k}_{i,j} \in \mathbb{R}^d$, and a value vector $\boldsymbol{v}_{i,j} \in \mathbb{R}^d$, $1 \le j \le m$, via the following operations:
\begin{equation}
%\begin{split}
\boldsymbol{Q}^{\text{S}}_i = \boldsymbol{W}^{\text{Q}}_\text{S} \boldsymbol{E}_i, 
\boldsymbol{K}^{\text{S}}_i = \boldsymbol{W}^{\text{K}}_\text{S} \boldsymbol{E}_i, 
\boldsymbol{V}^{\text{S}}_i = \boldsymbol{W}^{\text{V}}_\text{S} \boldsymbol{E}_i,
%\end{split}
\label{Eq_QKV}
\end{equation}
where $\boldsymbol{Q}^{\text{S}}_i = [ \boldsymbol{q}_{i,j} ]_{j = 1}^{m} \in \mathbb{R}^{d \times m}$, $\boldsymbol{K}^{\text{S}}_i = [ \boldsymbol{k}_{i,j} ]_{j = 1}^{m} \in \mathbb{R}^{d \times m}$, $\boldsymbol{V}^{\text{S}}_i = [ \boldsymbol{v}_{i,j} ]_{j = 1}^{m} \in \mathbb{R}^{d \times m}$, and $\boldsymbol{W}^{\text{Q}}_\text{S} \in \mathbb{R}^{d \times d}$, $\boldsymbol{W}^{\text{K}}_\text{S} \in \mathbb{R}^{d \times d}$, $\boldsymbol{W}^{\text{V}}_\text{S} \in \mathbb{R}^{d \times d}$ are the projection matrices that will be learned. Then we generate the attentional item embeddings $\widehat{\boldsymbol{e}}^u_{i,j} $ ($1 \le j \le m$) using the self-attention mechanism \cite{Vaswani2017}:
\begin{equation}
\widehat{\boldsymbol{E}}_i = \text{SelfAttention}(\boldsymbol{Q}^{\text{S}}_i, \boldsymbol{K}^{\text{S}}_i, \boldsymbol{V}^{\text{S}}_i) = \boldsymbol{V}^{\text{S}}_i \widehat{\boldsymbol{A}},
\label{Eq_Att}
\end{equation}
where $\widehat{\boldsymbol{E}}_i = [ \widehat{\boldsymbol{e}}^u_{i,j}]_{j = 1}^{m} \in \mathbb{R}^{d \times m}$, and 
%\begin{equation}
$\widehat{\boldsymbol{A}} = \text{softmax}\Big(\frac{({\boldsymbol{Q}^{\text{S}}_i})^{\text{T}} {\boldsymbol{K}^{\text{S}}_i}} {\sqrt{d}} \Big) \in \mathbb{R}^{m \times m}$
%\end{equation}
is the self-attention matrix. The cell $\widehat{a}_{j,l}$ at the $j$th row and $l$th column of $\widehat{\boldsymbol{A}}$ represents the attention score of the $j$th item $v^u_{i,j}$ to the $l$th item $v^u_{i,l}$ in session $\mathcal{S}^u_i$, and can be computed as
%\begin{equation}
$\widehat{a}_{j,l} = \frac{ \text{exp}(a_{j, l})} {\sum_{k = 1}^{m}a_{j, k}}$,
%\end{equation} 
where 
%\begin{equation}
$a_{j, l} = \frac{ {\boldsymbol{q}_{i,j}}^{\text{T}} \boldsymbol{k}_{i,l} } {\sqrt{d}}$ 
%\end{equation}
is the unnormalized attention score. Finally, the session embedding $\boldsymbol{s}^u_i$ is generated by summing up over the attentional item embeddings $\widehat{\boldsymbol{e}}^u_{i,j} $ ($1 \le j \le m$):
\begin{equation}
\boldsymbol{s}^u_i =\sum_{j=1}^{m} \widehat{\boldsymbol{e}}^u_{i,j}.
\label{Eq_SE}
\end{equation}
As we will see later, the last session embedding $\boldsymbol{s}^{u}_{T}$ will be used as the current short-term preference embedding $\boldsymbol{z}^{u}_{\text{short}}$ since it represents a user's current preference which might deviate from her/his long-term preference.

\subsubsection{Long-term Preference Learning}
The task of the long-term preference learning layer is to generate the long-term embedding $\boldsymbol{z}^u_{\text{long}} \in \mathbb{R}^d$ encoding the long-term preference of user $u$ at the global time scale, by using a multi-head self-attention module to fuse her/his local preferences represented by the session embeddings $\boldsymbol{s}^u_i \in \mathbb{R}^d$ ($1 \le i \le T$) which are outputs of the session embedding layer. As self-attention mechanism is not aware of the temporal positions of inputs, therefore to capture the temporal dependency between session embeddings, we first enhance each session embedding $\boldsymbol{s}^u_i$ by injecting a learnable position embedding $\boldsymbol{p}_i \in \mathbb{R}^d$ as follow:
\begin{equation}
\widehat{\boldsymbol{S}} = \boldsymbol{S} + \boldsymbol{P},
\end{equation}
where $\widehat{\boldsymbol{S}} = [\widehat{\boldsymbol{s}}^u_i]_{i=1}^{T} \in \mathbb{R}^{d \times T}$ is the enhanced session embedding matrix, $\boldsymbol{S} = [\boldsymbol{s}^u_i]_{i=1}^{T} \in \mathbb{R}^{d \times T}$ is the original session embedding matrix, $\boldsymbol{P} = [\boldsymbol{p}_i]_{i=1}^{T} \in \mathbb{R}^{d \times T}$ is the position embedding matrix, and $\widehat{\boldsymbol{s}}^u_i = \boldsymbol{s}^u_i + \boldsymbol{p}_i$.

Now we are going to generate the attentional session embedding matrix $\boldsymbol{Z} = [\boldsymbol{z}^u_i]_{i=1}^{T} \in \mathbb{R}^{d \times T}$, where each column $\boldsymbol{z}^u_i \in \mathbb{R}^{d}$ is the attentional session embedding corresponding to $\widehat{\boldsymbol{s}}^u_i$. We first define the basic multi-head self-attention function for matrix $\widehat{\boldsymbol{S}} = [\widehat{\boldsymbol{s}}^u_i]_{i=1}^{T} \in \mathbb{R}^{d \times T}$ as
\begin{equation}
\text{MultiHeadSelfAttention}(\widehat{\boldsymbol{S}}) = \boldsymbol{W}^{\text{O}} \text{Concat}( \boldsymbol{H}_1; \cdots; \boldsymbol{H}_h),
\label{Eq_MultiHeadAtt}
\end{equation}
where $\text{Concat}( \boldsymbol{H}_1; \cdots; \boldsymbol{H}_h) \in \mathbb{R}^{d \times T}$ represents the vertical concatenation of the $h$ heads $\boldsymbol{H}_j \in \mathbb{R}^{\frac{d}{h} \times T}$ ($1 \le j \le h$), and $\boldsymbol{W}^{\text{O}} \in \mathbb{R}^{d \times d}$ is a learnable projection matrix. Similar to Equation (\ref{Eq_QKV}), in order to generate each header $\boldsymbol{H}_j$, we build the query vector, the key vector, and the value vector for each enhanced session embedding via the following transformations:
\begin{equation}
%\begin{split}
\boldsymbol{Q}^{\text{H}}_j = \boldsymbol{W}^{\text{Q}}_{j} \widehat{\boldsymbol{S}}, 
\boldsymbol{K}^{\text{H}}_j = \boldsymbol{W}^{\text{K}}_{j} \widehat{\boldsymbol{S}}, 
\boldsymbol{V}^{\text{H}}_j = \boldsymbol{W}^{\text{V}}_{j} \widehat{\boldsymbol{S}}.
%\end{split}
\label{Eq_QKV_H}
\end{equation}
Then as same as Equation (\ref{Eq_Att}), each attention head $\boldsymbol{H}_j$ is obtained by the self-attention mechanism:
\begin{equation}
\begin{split}
\boldsymbol{H}_j &= \text{SelfAttention}(\boldsymbol{Q}^{\text{H}}_j, \boldsymbol{K}^{\text{H}}_j, \boldsymbol{V}^{\text{H}}_j) \\
&= \boldsymbol{V}^{\text{H}}_j \text{softmax}\Big( \frac{(\boldsymbol{Q}^{\text{H}}_j)^{\text{T}} \boldsymbol{K}^{\text{H}}_j} {\sqrt{d/h}}  \Big),
\end{split}
\end{equation} 
where $h$ is the number of heads and $\boldsymbol{W}^{\text{Q}}_{j}$, $\boldsymbol{W}^{\text{K}}_{j}$, $\boldsymbol{W}^{\text{V}}_{j} \in \mathbb{R}^{\frac{d}{h} \times d}$ are the learnable transformation matrices. Note that due to the sequential nature, it is supposed that the $i$th attentional session embedding $\boldsymbol{z}_i$ depends only on previous $i-1$ sessions. Therefore, we will mask the connection between $\boldsymbol{q}_i$ and $\boldsymbol{k}_j$ if $i < j$, by substituting zero for $\boldsymbol{q}_i^{\text{T}} \boldsymbol{k}_j$, where $\boldsymbol{q}_i$ and $\boldsymbol{k}_j$ are the $i$th column of $\boldsymbol{Q}$ (the query vector of $\boldsymbol{z}^u_i$) and the $j$th column of $\boldsymbol{K}$ (the key vector of $\boldsymbol{z}^u_j$), respectively.

As the transformer-based models do \cite{Vaswani2017}, to stabilize and accelerate the training of the multi-head self-attention network, we add a residual connection followed via a normalization: 
\begin{equation}
\text{Norm}\Big(\text{MultiHeadAttention}(\widehat{\boldsymbol{S}}) + \widehat{\boldsymbol{S}} \Big).
\label{Eq_MultiHeadAtt_Norm}
\end{equation}
For a matrix $\boldsymbol{X} = [\boldsymbol{x}_i]_{i=1}^{T} \in \mathbb{R}^{d \times T}$, the normalization function is defined as
\begin{equation}
%\begin{split}
\text{Norm}(\boldsymbol{X}) = [\text{norm}(\boldsymbol{x}_i)]_{i=1}^{T}, 
\text{norm}(\boldsymbol{x}) = \boldsymbol{\alpha} \otimes \frac{\boldsymbol{x} - \mu} {\sqrt{\sigma^2 + \epsilon}} + \boldsymbol{\beta},
%\end{split}
\end{equation}
where $\boldsymbol{\alpha} \in \mathbb{R}^d$ and $\boldsymbol{\beta} \in \mathbb{R}^d$ are learnable scaling factors and bias terms, $\otimes$ represents element-wise product, $\mu$ and $\sigma^2$ are the mean and variance of $\boldsymbol{x}$, respectively, and $\epsilon$ is a positive constant in case the illegal division incurred by zero variance.

%\begin{figure}[t]
%    \centering
%    \includegraphics[width=0.5\textwidth]{multi-head.pdf}
%    \caption{A multi-head self-attention block.}
%    \label{fig:multihead}
%\end{figure}

To capture non-linear interactions between the latent dimensions, we further apply a feed-forward network FFN to the output of Equation (\ref{Eq_MultiHeadAtt_Norm}). Then the final attentional session embedding matrix can be obtained by:
\begin{equation}
\boldsymbol{Z} = \text{FFN} \Bigg( \text{Norm}\Big(\text{MultiHeadAttention}(\widehat{\boldsymbol{S}}) + \widehat{\boldsymbol{S}} \Big)  \Bigg),
\label{Eq_Z}
\end{equation}
where the feed-forward network is defined as
\begin{equation}
\begin{split}
 \text{FFN}(\boldsymbol{X} &= [\boldsymbol{x}_i]_{i=1}^{T} \in \mathbb{R}^{d \times T}) = [\text{ffn}(\boldsymbol{x}_i)]_{i=1}^{T}, \\
 \text{ffn}(\boldsymbol{x} \in \mathbb{R}^d) &= \boldsymbol{W}^{\text{F}}_2 \max(0, \boldsymbol{W}^{\text{F}}_1 \boldsymbol{x} + \boldsymbol{b}_1) + \boldsymbol{b}_2,
\end{split}
\end{equation}
where $\boldsymbol{W}^{\text{F}}_1 \in \mathbb{R}^{4d \times d}$, $\boldsymbol{W}^{\text{F}}_2 \in \mathbb{R}^{d \times 4d}$, $\boldsymbol{b}_1 \in \mathbb{R}^{4d}$, and $\boldsymbol{b}_2 \in \mathbb{R}^{d}$ are the learnable transformation matrices and bias terms. Note that Equation (\ref{Eq_Z}) constitutes a multi-head self-attention block, and in practice we can stack more than one multi-head self-attention block (i.e., iteratively apply Equation (\ref{Eq_Z})) to enhance the expressiveness of the attentional session embeddings.

Once the attentional session embeddings $\boldsymbol{z}^u_i$ are obtained through Equation (\ref{Eq_Z}), we can finally generate the long-term preference embedding $\boldsymbol{z}^u_{\text{long}}$ via a vanilla attention module:
\begin{equation}
%\begin{split}
\boldsymbol{z}^u_{\text{long}} = \sum_{i=1}^{T} \omega_{i} \boldsymbol{z}^u_i, 
\omega_{i} = \frac{ \exp\Big(\boldsymbol{f}^{\text{T}}_u \phi(\boldsymbol{W}_{\text{L}} \boldsymbol{z}^u_{i} + \boldsymbol{b}_{\text{L}}) \Big) } {\sum_{i=1}^{T} \exp\Big(\boldsymbol{f}^{\text{T}}_u \phi(\boldsymbol{W}_{\text{L}} \boldsymbol{z}^u_{i} + \boldsymbol{b}_{\text{L}}) \Big)},
%\end{split}
\label{Eq_LE}
\end{equation}
where $\omega_{i}$ is the attentional coefficient of $\boldsymbol{z}^u_{i}$, $\phi$ is the ReLU activation function, $\boldsymbol{W}_{\text{L}} \in \mathbb{R}^{d \times d}$ and $\boldsymbol{b}_{\text{L}} \in \mathbb{R}^d$ are the learned transformation matrix and bias terms, respectively, and $\boldsymbol{f}_u \in \mathbb{R}^{d}$ is the embedding of user $u$. Similar to item embedding, $\boldsymbol{f}_u$ is also obtained by a lookup over a learnable user embedding matrix $\boldsymbol{W}^{\text{U}} \in \mathbb{R}^{d \times N}$:
%\begin{equation}
$\boldsymbol{f}_u = \boldsymbol{W}^{\text{U}} \boldsymbol{u}$,
%\label{Eq_UE}
%\end{equation}
where $\boldsymbol{u} \in \mathbb{R}^{N}$ is the one-hot encoding of user $u$.

\subsection{Neural Time Gate}

Now the long-term preference embedding $\boldsymbol{z}^u_{\text{long}}$ and the short-term preference embedding $\boldsymbol{z}^u_{\text{short}}$ have been prepared by the hierarchical self-attention network. Next we will produce the final preference embedding used for rating prediction, by fusing $\boldsymbol{z}^u_{\text{long}}$ and $\boldsymbol{z}^u_{\text{short}}$ via the proposed neural time gate.

The task of the neural time gate is to adjust the contributions of the long-term preference embedding and the current short-term preference embedding at dimension level, based on the lag $\Delta t$ between the time of last interaction and the time when a recommendation needs to be made. To encode the time lag into an intermediate embedding, we first discretize it by its multiples of the minimum time gap $\Delta_{\text{min}}$ bewteen any two successive interactions of a give user. In this idea, the discretized time lag $\delta \in \mathbb{N}$ is computed as 
\begin{equation}
\delta = \min(\lceil \frac{\Delta t}{\Delta_{\text{min}}} \rceil, C),
\label{Eq_TimeInterval}
\end{equation}
where $C \in \mathbb{N}$ represents the maximal value of $\delta$. By Equation (\ref{Eq_TimeInterval}), the $\Delta t$ is mapped to a positive number not more than $C$. Then we can get the time embedding $\boldsymbol{y} \in \mathbb{R}^d$ by a lookup over a learnable embedding matrix $\boldsymbol{Y} \in \mathbb{R}^{d \times C}$ as follow:
\begin{equation}
\boldsymbol{y} = \boldsymbol{Y} \boldsymbol{\delta},
\label{Eq_Y}
\end{equation}
where $\boldsymbol{\delta} \in \mathbb{R}^{C}$ is the one-hot vector of the discretized time lag. Then the normalized gating vector $\boldsymbol{g} \in \mathbb{R}^{d}$ can be computed by the Sigmoid function
\begin{equation}
\boldsymbol{g} = \text{sigmoid}(\boldsymbol{W}_{l} \boldsymbol{z}^u_{\text{long}} + \boldsymbol{W}_{s} \boldsymbol{z}^u_{\text{short}} +  \boldsymbol{W}_{\delta} \boldsymbol{y} + \boldsymbol{b}_g),
\label{Eq_G}
\end{equation}
where $\boldsymbol{W}_{l}$, $\boldsymbol{W}_{s}$, $\boldsymbol{W}_{\delta} \in \mathbb{R}^{d \times d}$ and $\boldsymbol{b}_g \in \mathbb{R}^d$ are the learnable weight matrices and bias vector, respectively. Finally, the fused preference embedding of the given user $u$ is obtained by the following fusion based on $\boldsymbol{g}$:
\begin{equation}
\boldsymbol{z}_u = \boldsymbol{g} \otimes  \boldsymbol{z}^u_{\text{short}} + (\boldsymbol{1} -  \boldsymbol{g}) \otimes  \boldsymbol{z}^u_{\text{long}},
\label{Eq_Fusion}
\end{equation}
where $\otimes$ represents element-wise product. Note that $\boldsymbol{g}$ is a vector rather than a scalar, which enables the neural time gate to regulate the contributions of the long-term preference and short-term preference at the dimension granularity.

\subsection{Rating Prediction}
Finally, we adopt a dot product of the fused preference embedding $\boldsymbol{z}_u$ and the item embedding $\boldsymbol{e}_v$ as the prediction of the normalized rating that user $u$ gives to item $v$, i.e.,
\begin{equation}
\widehat{r}_{u, v} = \text{sigmoid}(\boldsymbol{z}^{\text{T}}_u \boldsymbol{e}_v).
\label{Eq_Output}
\end{equation}

\subsection{Model Learning}
\subsubsection{Training Set Building}
We first build a training set $\mathcal{O}^{u}$ for each user $u$, where each instance $o \in \mathcal{O}^{u}$ is a sequence of $T+1$ sessions, i.e $o = \langle \mathcal{S}^u_1, \cdots, \mathcal{S}^u_T, \mathcal{S}^u_{T+1} \rangle$. During the training, the first $T$ sessions $\mathcal{S}^u_1, \cdots, \mathcal{S}^u_T$ are used as input of the model, while the last session $\mathcal{S}^u_{T+1}$ serves as ground truth for the supervision of the training. For a training data set, a user's sessions are divided whenever the time interval between two successive interactions is more than a chosen threshold. %As we will see in the experiments, for a dataset we will investigate the distribution of the different time intervals, and choose as the threshold the one that dominates the distribution.

Since the length of different session sequences might be different, for a sequence with length greater than $T$, we use a sliding window of width $T$ to split it into subsequences of the fixed length $T+1$, while for a sequence with length less than or equal to $T$, we use the first session to pad to the left to the sequence until its length becomes $T+1$. Similarly, the length of a session (i.e. the number of items contained in a session) might also be different from each other. For a session whose length is less than the length of the longest session, we will repeatedly pad that session with its last item until its length becomes $m$, where $ m =\max_{u \in \mathcal{U}, 1 \le i \le T } (|\mathcal{S}^u_i|) $.
\subsubsection{Model Optimization}

As our goal is to recommend a ranked list of items, we are more interested in the relative ranking order of the rating predictions rather than their absolute values. For a training instance $o = \langle \mathcal{S}_1, \cdots, \mathcal{S}_T, $ $\mathcal{S}_{T+1} \rangle$, let $\mathcal{V}_o^{+} = \mathcal{S}_{T+1}$ be the ground truth. For each item $v \in \mathcal{V}_o^{+}$, we sample an unobserved item $v' \notin \mathcal{V}_o^{+}$ to form a negative sample set $\mathcal{V}_o^{-}$. We expect the predicted rating of an item $v \in \mathcal{V}_o^{+}$ will be greater than that of an item $v' \in  \mathcal{V}_o^{-}$, i.e., $\widehat{r}_{u, v} > \widehat{r}_{u, v'}$. For this purpose, we define a pair-wise loss function based on the principle of Bayesian Personalized Ranking (BPR) \cite{Rendle2009}:
\begin{equation}
L(\boldsymbol{\Theta}) = - \sum_{u \in \mathcal{U}} \sum_{o \in \mathcal{O}^u} \sum_{v \in \mathcal{V}_o^{+}, v' \in \mathcal{V}_o^{-}} \log \sigma (\widehat{r}_{u, v} - \widehat{r}_{u, v'}) + \lambda \parallel \boldsymbol{\Theta} \parallel_2^2,
\end{equation}
where $\Theta$ represents all the learnable parameters and $\lambda$ is a nonnegative parameter controlling the contribution of the regularization term. In the experiments, we will apply Adam algorithm \cite{Diederik2015} to optimize our model.

\begin{table*}[t]
\centering
\footnotesize
\caption{The statistics of datasets}
\label{tab:statistics}
\begin{tabular}{p{1.9cm}|p{1.0cm}<{\centering}p{0.8cm}<{\centering}p{1.2cm}<{\centering}p{1.5cm}<{\centering}p{1.5cm}<{\centering}p{1.2cm}<{\centering}p{1.0cm}<{\centering}}
\toprule
Dataset&\#Users &\#Items &\#Interactions &Avg. \#sessions per user &Avg. length per session &Density\\
\hline
Amazon Book       & 4,621  & 170,474  & 517,556  & 12  & 9 & 0.0006 \\
Amazon Video      & 1,709  & 12,434  & 64,298   & 8   &5  & 0.0030 \\
Movielens-1M      & 5,492  & 3,692   & 970,346  & 29  & 6 & 0.0478 \\
Lastfm        	 & 953    &22,372   &16,641,736 & 950 & 18 &0.7805 \\
\hline
\end{tabular}
\end{table*}

\section{Experiments}
%The experiments mainly aim to answer the following research questions:
%\begin{enumerate}[\textbf{RQ}1]
%\item How does TLSRec perform as compared with state-of-the-art sequential recommendation methods?
%\item How do the different components of TLSRec affect its performance?
%\item How can the effectiveness of TLSRec be illustrated with intuitive and visualizable case studies?
%\item How do the hyper-parameters, including the number $T$ of input sessions, the latent dimensionality $d$, and the head amount $h$ of the multi-head self-attention in the long-term preference learning layer, influence the performance of TLSRec?
%\item How does the running time of TLSRec vary with different input sizes.
%\end{enumerate}
%In the following, we will first introduce the experimental setting and then analyze the detailed results of the experiments.

\subsection{Experimental Setting}

\subsubsection{Datasets}

We conduct the experiments on four real-world datasets whose statistics are summarized in Table \ref{tab:statistics}. In this paper, we only consider implicit feedbacks (e.g. clicks), and hence explicit feedbacks (e.g. ratings) in datasets are simply regarded as implicit interactions. As has mentioned before, in order to split an interaction sequence into sessions, on each dataset we will investigate the distribution of the time gaps between any two successive interactions, and choose as the threshold the time gap that accounts for the most part of the distribution. On each dataset, we randomly select 70\%of the data as training set, 10\% as validation set, and the remaining 20\% as testing set, and repeat such procedure 10 times and report the average results.

\begin{itemize} 
\item \textbf{Amazon Book} Amazon Book is a dataset collected from Amazon, which contains 517,556 ratings to 170,474 books given by 4,621 users. By investigating the distribution of the time gaps between any two successive interactions, we find that most time gaps are less than 2 days. Therefore, in Amazon Book, we split a historical interaction sequence into sessions whenever the time interval between two successive interactions is more than 2 days. 

\item \textbf{Amazon Video} Amazon Video is another dataset collected from Amazon, which contains 64,298 ratings to 12,434 videos given by 1,709 users. In Amazon Video, the sessions are extracted using the same time gap threshold as in Amazon Book.

\item \textbf{MovieLens-1M} MovieLens-1M is a user-movie dataset collected from MovieLens website, which contains 970,346 ratings to 3,692 movies given by 5,492 users. For MovieLens-1M, the time gap threshold is set to 2 hours with the same method as for the Amazon datasets. 

\item \textbf{Lastfm} Lastfm is a freely-available collection of audio features and metadata for a million contemporary popular music tracks \cite{Bertin-Mahieux2011}, consisting of tuples of user, timestamp, artist, and song listened to. As Lastfm contains an overwhelming amount of songs, which causes an expensive requirement of huge amount of memory, we treat the artists instead of the songs as items, with the same approach as taken by \cite{Ruocco2017}, and we obtain 16,641,736 interactions of 953 users with 22,372 artists. Finally, we split an interaction sequence into sessions for Lastfm using the same time gap threshold as for MovieLens-1M.

\end{itemize}

\subsubsection{Baseline Methods}
%We compare TLSRec with following ten state-of-the-art methods for sequential recommendation:
We compare TLSRec with ten state-of-the-art methods for sequential recommendation, including two RNN based models (DREAM and II-RNN), six attention based models (NARM, ANAM, SHAN, SASRec, BERT4Rec, and TiSASRec), and two GNN based models (SURGE and RetaGNN).

\begin{itemize} 

%\item \textbf{HHMM }\cite{HosseinzadehAghdam2015} HHMM uses a hierarchical hidden Markov model to characterize the dynamics of user preference, where user preference is inferred by the hidden variables at upper layer with maximizing the likelihood of the observed transition sequence of the states at the lower layer.

\item \textbf{DREAM }\cite{Yu2016} DREAM is an RNN based model for next basket recommendation, which not only learns a dynamic representation for a user but also captures global sequential features among baskets (sessions) to gain a comprehensive understanding of users' purchase interests and consequently recommend items that each user most probably purchase in the next visit.

\item \textbf{II-RNN} \cite{Ruocco2017} II-RNN is a hierarchical RNN model for sequential recommendation, which not only models a user's short-term preference by an intra-session RNN layer, but introduces an inter-session RNN layer to capture the dependency between sessions as well.

\item \textbf{NARM} \cite{Li2017} NARM is an attention based model for session-based recommendation, which uses a hybrid encoder with an attention mechanism to model the user’s sequential behavior and capture users' intent in the current session.

\item \textbf{ANAM} \cite{Ting2018} ANAM is an attribute-aware model for next basket (session) recommendation, which adopts an attention mechanism to explicitly model user’s evolving preference for items, and utilizes a hierarchical architecture to incorporate the attribute information of items.

\item \textbf{SHAN} \cite{Ying2018} SHAN is a sequential recommendation model based on a two-layer hierarchical attention network, where the first attention layer learns user long-term preferences based on the historical purchased items, while the second one generates user's final representation by fusing the user's long-term preference and short-term preference.

\item \textbf{SASRec} \cite{Kang2018} SASRec uses a self-attention mechanism combined with position embeddings to capture the semantics of user's long-term preference, which can identify relevant items by adaptively assigning weights to previous items at each time step. 

\item \textbf{BERT4Rec} \cite{Sun2019} BERT4Rec employs a deep bidirectional self-attention mechanism to model user behavior sequences, with the optimization objective to predicting the random masked items in the sequence by jointly conditioning on their left and right context.

\item \textbf{TiSASRec} \cite{Li2020}: TiSASRec is a time interval aware model for sequential recommendation, which incorporates the information of the relative time interval between any two items into a self-attention mechanism to weight the different items during the learning of user preference.

\item \textbf{SURGE} \cite{chang2021sequential}: SURGE is a GNN-based model which integrates implicit feedbacks with explicit ones in long-term user behaviors into clusters in the graph by re-constructing loose item sequences into tight item-item interest graphs based on metric learning.

\item \textbf{DHCN} \cite{xia2021self}: DHCN models session-based data as a hypergraph and captures the high-order relations among items and the cross-session information with a dual channel hypergraph convolutional network.

\end{itemize}

\subsubsection{Evaluation Metrics}
We choose the widely used Hit rate and MAP (Mean Absolute Precision) to evaluate the performance of TLSRec. Let $\mathcal{S}_u$ and $\widehat{\mathcal{S}}_u$ be the ground truth session and the set of the predicted top-$k$ ranked items, and then the Hit rate and MAP can be defined as follows:

\begin{equation}
\text{Hit}@k = \frac{1}{ |\mathcal{U} |}\sum_{u \in \mathcal{U} } \mathbb{I} \big( | \mathcal{S}_u \cap \widehat{\mathcal{S}}_u | > 0 \big),
\end{equation}

\begin{equation}
\text{MAP}@k =  \frac{1}{|\mathcal {U} |} \bigg( \sum_{u \in \mathcal{U}}\sum_{i \in \mathcal{S}_u \cap \widehat{\mathcal{S}}_u } \frac{ \sum_{j \in \mathcal{S}_u \cap \widehat{\mathcal{S}}_u } \mathbb{I}(\gamma_{uj} < \gamma_{ui})+1} {\gamma_{ui}} \bigg),
\end{equation}
where $|\widehat{\mathcal{S}}_u| = k$, $\mathbb{I}(x) = 1$ if $x$ is true, otherwise $\mathbb{I}(x) = 0$, and $\gamma_{ui}$ is the predicted rank of item $i$ for user $u$.

\subsubsection{Hyper-parameter Setting}
To divide interactions into sessions, we set the time gap threshold $\epsilon$ to $48$ hours for Amazon Book and Amazon Video, and $2$ hours for MovieLens-1M and Lastfm. To build the training and testing instances, we set the number $T$ of sessions in an instance to $10$, $8$, $6$, and $4$ for Amazon Book, Amazon Video, Movielens-1M, and Lastfm, respectively. The embedding dimensionality $d$ is set to $64$ for Amazon Book, Movielens-1M, and Lastfm, and $32$ for Amazon Video. At last, the number of heads in multi-head self-attention network in long-term preference learning layer is set to $8$ for all datasets. On all datasets, the learning rate, batch size, and the dropout rate are set to 0.001, 128, and 0.5, respectively. At the same time, the optimal hyper-parameters of baselines are fine-tuned on validation sets. %The hyper-parameter settings are shown in Table \ref{tab:hyperparameters}. 

\subsection{Performance Comparison with Baselines}

%The results of the comparison of TLSRec and the baselines are shown in Tables \ref{tab:Hit_Comparison} and \ref{tab:MAP_Comparison}.

\begin{table}[!t]
\renewcommand\tabcolsep{2.0pt}
\centering
\footnotesize
\caption{Hit$@k$ comparison with baselines}
\centering
\label{tab:Hit_Comparison}
\begin{tabular}{lcccccccc}
\toprule  
Methods& \multicolumn{2}{c}{Amazon Book} &\multicolumn{2}{c}{Amazon Video} & \multicolumn{2}{c}{Movielens-1M}&\multicolumn{2}{c}{Lastfm}\\
\cmidrule(r){2-3} \cmidrule(r){4-5} \cmidrule(r){6-7} \cmidrule(r){8-9}
&$k=20$ &$k=30$ &$k=20$ &$k=30$ &$k=20$ &$k=30$ &$k=20$ &$k=30$\\
\hline
%HHMM &0.0335 &0.0453 &0.0358 &0.0430 &0.3252 &0.4131 &0.5875 &0.6543\\

DREAM &0.0440 &0.0609 &0.0553 &0.0644 &0.3837 &0.4713 &0.7323 &0.7861 \\

II-RNN &0.0668 &0.0907 &0.0821 &0.0966 &0.4902 &0.5854 &0.6311 &0.6864 \\
\hline

NARM& 0.0634 &0.0838 &0.0339 &0.0449 &0.4208 &0.5066 &0.7097 &0.7585 \\

ANAM  &0.0651  &0.0971  &0.0819 &0.1019 &0.3381 &0.4123 &0.7287&0.7991 \\

SHAN  &0.0480 &0.0632 &0.0627 &0.0899 &0.3337 &0.4237 &0.5556 &0.6100 \\

SASRec  &0.1209&0.1663 &0.1884 &0.2458 &0.4476 &0.5413 &\underline{0.8065} &\underline{0.8260}\\

BERT4Rec  &0.0902&0.1335 &0.1744 &0.2167 &0.3917 &0.4922 &0.6515 &0.7748\\

TiSASRec &0.1842 &0.2356 &0.2133 &0.2680 &0.5025 &0.5850 &0.7630 &0.8253\\

\hline

SURGE  &\underline{0.2219}&\underline{0.3103} &0.1961 &0.2048 &0.5077 &0.5110 &0.7915 &0.8060\\

DHCN  &0.1991&0.2736 &\underline{0.2308} &\underline{0.2780} &\underline{0.5301} &\underline{0.5922} &0.7551 &0.8180\\

\hline

TLSRec &\textbf{0.3438} &\textbf{0.4161} &\textbf{0.2423}&\textbf{0.2884}&\textbf{0.5640} &\textbf{0.6522}&\textbf{0.8086}&\textbf{0.8439}\\
\bottomrule
\end{tabular}
\end{table}

\begin{table}[!t]
\renewcommand\tabcolsep{2.0pt}
\centering
\footnotesize
\caption{MAP$@k$ comparison with baselines}
\centering
\label{tab:MAP_Comparison}
\begin{tabular}{lcccccccc}
\toprule  
Methods& \multicolumn{2}{c}{Amazon Book} &\multicolumn{2}{c}{Amazon Video} & \multicolumn{2}{c}{Movielens-1M}&\multicolumn{2}{c}{Lastfm}\\
\cmidrule(r){2-3} \cmidrule(r){4-5} \cmidrule(r){6-7} \cmidrule(r){8-9}

&$k=20$ &$k=30$ &$k=20$ &$k=30$ &$k=20$ &$k=30$ &$k=20$ &$k=30$\\
\hline
%HHMM &0.0016 &0.0017  & 0.0021 &0.0024 &0.0182 &0.0216  &0.0550 &0.0609 \\
%\hline
DREAM &0.0020 &0.0021  & 0.0027 &0.0028 &0.0214 &0.0230  &0.0661 &0.0769 \\

II-RNN & 0.0025 &0.0035 &0.0031 &0.0038 &0.0271  &0.0311 &0.0899 & 0.0928  \\
\hline

NARM  &0.0031 &0.0033 &0.0023 &0.0028 & 0.0185 & 0.0214 & 0.0731 &0.0784 \\

ANAM  &0.0021 &0.0031 &0.0038 &0.0047  &0.0184 &0.0214 &0.0674 &0.0713 \\

SHAN  &0.0018 &0.0019 &0.0061 &0.0066  &0.0188 &0.0214 &0.0690 &0.0700 \\

SASRec  &0.0055 &0.0062 &0.0145 &0.0163 &0.0352 &0.0436 &0.0892 &0.0966\\

BERT4Rec  &0.0033&0.0047 &0.0092 &0.0166 &0.0279 &0.0402 &0.0755 &0.0834\\

TiSASRec  &0.0062 &0.0073 &\underline{0.0155} &\underline{0.0172} &\underline{0.0372} &\underline{0.0437} &\underline{0.0981} &\underline{0.1046} \\
\hline

SURGE  &\underline{0.0091}&\underline{0.0104} &0.0096 &0.0138 &0.0366 &0.0411 &0.0905 &0.1006\\

DHCN  &0.0075&0.0776 &0.0082 &0.0170 &0.0233 &0.0392 &0.0808 &0.0984\\
\hline

TLSRec  &\textbf{0.0130} &\textbf{0.0144} &\textbf{0.0167}&\textbf{0.0175} &\textbf{0.0381} &\textbf{0.0438} &\textbf{0.1044}&\textbf{0.1097}\\
\bottomrule
\end{tabular}
\end{table}

The results of the comparison with baseline methods are shown in Tables \ref{tab:Hit_Comparison} and \ref{tab:MAP_Comparison}. It can be seen that TLSRec outperforms the two RNN based models, DREAM, and II-RNN, in terms of Hit$@k$ and MAP$@k$. The RNN based models often suffer from the problem of long-term dependency, which causes they prefer to memorize the preference more reflected by recent sessions than by distant sessions. As the recent sessions dominate the learning of user preference, the RNN based models are more susceptible to the fluctuates of users' short-term preference and cannot sufficiently capture the long-term preference that is more stable. In contrast to the RNN based models, TLSRec learns an embedding for long-term preference by pooling local preference embeddings of sessions with a hierarchical self-attention network, which enables TLSRec to perceive the long-term dependency between sessions and smooth out the preference fluctuates, and consequently better understand the stable long-term preference of a user. %At the same time, TLSRec uses a neural time gate to adaptively regulate the contributions of the long-term preference and short-term preference to a user's current preference, which helps TLSRec to capture users' real preference more comprehensively at the time when a recommendation needs to be made. 

We also observe that TLSRec outperforms the six attention based models. Essentially, NARM, ANAM, SASRec, and TiSASRec only learn the short-term preference of a user by fusing the embeddings of the recent interactions with an attention network, which lack the knowledge about the user's long-term preference and consequently tend to be hindered by the fluctuates of the user's preference. In contrast, TLSRec can learn not only the short-term preference but the long-term preference as well, particularly with a hierarchical self-attention network which makes it able to capture the dependency between sessions. Although SHAN and BERT4Rec consider both long-term preference and short-term preference, however it learns the current preference by fusing them with scalar coefficients produced by an attention mechanism, which makes it unable to weight the contributions of the two preferences at a finer granularity level like TLSRec does.

It can be also observed that GNN based models are inferior to TLSRec. Although the GNN based models can capture the high-order interactions between items among sessions, they often essentially focus on the learning of the current session without differentiating the impacts of the long-term and short-term preferences

Unlike the baseline methods, TLSRec uses a gate vector produced by a neural time gate based on the time distance to fuse the long-term preference and short-term preference embeddings, which benefits the learning of the current preference from two perspectives. First, the contributions of the long-term preference and short-term preference are reasonably regulated by the distance to current time, and the shorter it is, the more proportion the short-term preference accounts for. Second, the dimensions of the gate vector play the role weighting the dimensions of the preference embeddings during their fusion, which offers a finer-grained fusion.
%makes TLSRec be able to control the fusion at the finer level.

\subsection{Ablation Experiments}

%\begin{table*}[t]
%\renewcommand\tabcolsep{1.4pt}
%\footnotesize
%\caption{Summary of the variants of TLSRec}
%\centering
%\label{tab:variants}
%\begin{tabular}{p{2.5cm}<{\centering}|p{2.5cm}<{\centering}p{2.5cm}<{\centering}p{2.0cm}<{\centering}p{2.0cm}<{\centering}}
%\toprule 
% Variants&Self-attention for short-term preference learning&Self-attention for long-term preference learning & Multi-head self-attention &Neural Time Gate\\
%\hline
%TLSRec-S &$\times$ & $\surd$ &$\surd$  & $\surd$ \\
%TLSRec-L & $\surd$ &$\times$  & $\surd$  & $\surd$  \\
%TLSRec-M &$\surd$ &$\surd$ & $\times$ & $\surd$ \\
%TLSRec-G+A &$\surd$ & $\surd$ &$\surd$ &$\times$ \\
%TLSRec-G+S &$\surd$ & $\surd$ &$\surd$ & $\times$ \\ %& replaced by self-attention \\
%TLSRec-G+M &$\surd$ & $\surd$ &$\surd$ & $\times$ \\ %replaced by multi-head attention \\
% \hline
%\end{tabular}
%\end{table*}

Now we investigate the effectiveness of the hierarchical self-attention network, the neural time gate, and the multi-head self-attention mechanism in the long-term preference learning. For this purpose, we will compare TLSRec with its variants as follows:

\begin{itemize} 

\item{\textbf{TLSRec-S}} Compared with TLSRec, TLSRec-S removes the self-attention network before the average pooling function for the generation of session embeddings in short-term preference learning layer.

\item{\textbf{TLSRec-L}} Symmetrically, TLSRec-L removes self-attention network before the vanilla attention network for the generation of the long-term preference embedding.

\item{\textbf{TLSRec-M}} Compared with TLSRec, TLSRec-M replaces the multi-head self-attention network with a single-head self-attention network in the long-term preference learning layer.

\item{\textbf{TLSRec-G+A}} TLSRec-G+A is a variant of TLSRec where the neural time gate is replaced with a pooling function which generates the current preference embedding $\boldsymbol{z}_u$ by averaging the long-term preference embedding and the short-term preference embedding.

\item{\textbf{TLSRec-G+S}} TLSRec-G+S is a variant of TLSRec where the neural time gate is replaced with a self-attention mechanism. TLSRec-G+S first generates the attentional long-term embedding and short-term embedding with attention to each other, and then generates the current preference embedding $\boldsymbol{z}_u$ with the sum of them.

\item{\textbf{TLSRec-G+M}} TLSRec-G+M is a variant of TLSRec where the neural time gate is replaced with a multi-head attention mechanism. TLSRec-G+M generates the current preference embedding similarly to TLSRec-G+S, with the exception of using the multi-head attention to generate the attentional long-term and short-term embeddings. In TLSRec-G+M, the number of attention heads is set to the same value in TLSRec. 

\end{itemize}

\begin{table}[!t]
\renewcommand\tabcolsep{2.0pt}
\centering
\footnotesize
\caption{Hit$@k$ comparison with the variants}
\centering
\label{tab:ablation_Hit}
\begin{tabular}{lcccccccc}
\toprule  
Methods& \multicolumn{2}{c}{Amazon Book} &\multicolumn{2}{c}{Amazon Video} & \multicolumn{2}{c}{Movielens-1M}&\multicolumn{2}{c}{Lastfm}\\
\cmidrule(r){2-3} \cmidrule(r){4-5} \cmidrule(r){6-7} \cmidrule(r){8-9}

&$k=20$ &$k=30$ &$k=20$ &$k=30$ &$k=20$ &$k=30$ &$k=20$ &$k=30$\\
\hline

TLSRec-S &0.2741  &0.3530  &0.1567 &0.2073 &0.5111 &0.6057 &0.7604 &0.8064 \\

TLSRec-L &0.3034 &0.3692 &0.1789 &0.2221 &0.5466 &0.6319 &0.7679& 0.8160\\

TLSRec-M &0.3039 &0.3736 &0.1875 &0.2344 &0.5611 &0.6495 &0.8060 &0.8435\\

TLSRec-G+A  &0.2989  &0.3677  &0.1702 &0.2134 &0.5317 &0.6263 &0.8053 &0.8427 \\

TLSRec-G+S  &0.3070  &0.3712  &0.1770 &0.2315 &0.5534 &0.6476 &0.8035 &0.8427 \\

TLSRec-G+M  &0.3213  &0.3894  &0.1965 &0.2448 &0.5634 &0.6514 &0.8065 &0.8436 \\

TLSRec &\textbf{0.3438} &\textbf{0.4161} &\textbf{0.2423}&\textbf{0.2884}&\textbf{0.5640} &\textbf{0.6522}&\textbf{0.8086}&\textbf{0.8439}\\
\bottomrule
\end{tabular}
\end{table}

\begin{table}[t]
\renewcommand\tabcolsep{2.0pt}
\centering
\footnotesize
\caption{MAP$@k$ comparison with the variants}
\centering
\label{tab:ablation_MAP}
\begin{tabular}{lcccccccc}
\toprule  
Methods& \multicolumn{2}{c}{Amazon Book} &\multicolumn{2}{c}{Amazon Video} & \multicolumn{2}{c}{Movielens-1M}&\multicolumn{2}{c}{Lastfm}\\
\cmidrule(r){2-3} \cmidrule(r){4-5} \cmidrule(r){6-7} \cmidrule(r){8-9}

&$k=20$ &$k=30$ &$k=20$ &$k=30$ &$k=20$ &$k=30$ &$k=20$ &$k=30$\\
\hline

TLSRec-S &0.0077  &0.0087  &0.0119  &0.0127 &0.0245 &0.0300&0.0928 &0.0975 \\

TLSRec-L &0.0113 &0.0130 &0.0130 &0.0139 &0.0367 &0.0421 &0.0822 &0.0866 \\

TLSRec-M &{0.0117}  &{0.0131} &{0.0159} &{0.0169 }&{0.0372} &{0.0428} &0.0962 &0.1018 \\

TLSRec-G+A  &0.0110  &0.0121  &0.0149 &0.0157 &0.0360 &0.0416 &{0.1016} &{0.1078} \\

TLSRec-G+S  &{0.0110}  &{0.0121} &{0.0159} &{0.0166 }&{0.0362} &{0.0418} &0.1016 &0.1077 \\

TLSRec-G+M  &{0.0119}  &{0.0133} &{0.0163} &{0.0171 }&{0.0376} &{0.0430} &0.1028 &0.1094 \\

TLSRec  &\textbf{0.0130} &\textbf{0.0144} &\textbf{0.0167}&\textbf{0.0175} &\textbf{0.0381} &\textbf{0.0438} &\textbf{0.1044}&\textbf{0.1097}\\
\bottomrule
\end{tabular}
\end{table}

%These variants are summarized in Table \ref{tab:variants}, and 

The results are shown in Tables \ref{tab:ablation_Hit} and \ref{tab:ablation_MAP}. We can see that compared with TLSRec, the performance of each variant significantly declines, which verifies the effectiveness of different components of TLSRec. Particularly, the comparison between TLSRec, TLSRec-S and TLSRec-L shows the performance gain incurred by the proposed hierarchical self-attention network, which verifies its ability to better capture the dependency between items for the short-term preference learning and the dependency between sessions for the long-term preference learning. At the same time, we can also note that TLSRec performs much better than TLSRec-M, which is due to the ability of the multiple attention heads to enhance the long-term preference learning by perceiving the finer-grained interactions between dimensions of session preference embeddings. At last, we see that compared with TLSRec-G+A, TLSRec-G+S and TLSRec-G+M, the performance of TLSRec is remarkably improved because of the advantages of the proposed neural time gate. First, the results demonstrate the effectiveness of using the time lag aware gate vector to adaptively regulate the contributions of the long-term preference and short-term preference for the learning of the current preference. Second, the results also show that the fine-grained fusion of the long-term and short-term embeddings with dimension-wise weights offered by the neural time gate is superior to the coarse fusion with manually predefined vector-wise weights.

%First, it verifies the effectiveness of using the time lag between the last interaction and the time of making a recommendation to regulate the contributions of the long-term preference and the short-term preference for the learning of the current preference. Second, the results also shows the effectiveness of using a gate vector to fuse the long-term and short-term preference embeddings at the finer-grained dimension level. In addition, we note that TLSRec-G+S and TLSRec-G+A perform very similarly because they are both scalar based weighting schema and TLSRec-G+A is a specialization of TLSRec-G+S. TLSRec-G+M is the best of all variants. It proves that fine-grained tradeoff of the long-term and short-term preferences is helpful, and TLSRec has a significant improvement on the Amazon datasets with sparse interactive data compared with TLSRec-G+M. It shows the effectiveness of our proposed neural time gate in helping users with infrequent interactions capture accurate current preference for sequential recommendation.

\subsection{Case Study}
\begin{figure}[!t]
\centering
\begin{minipage}{0.2\textwidth}
\centering
\subfigure[User ID '237']{\label{a}\includegraphics[width=1.0\textwidth]{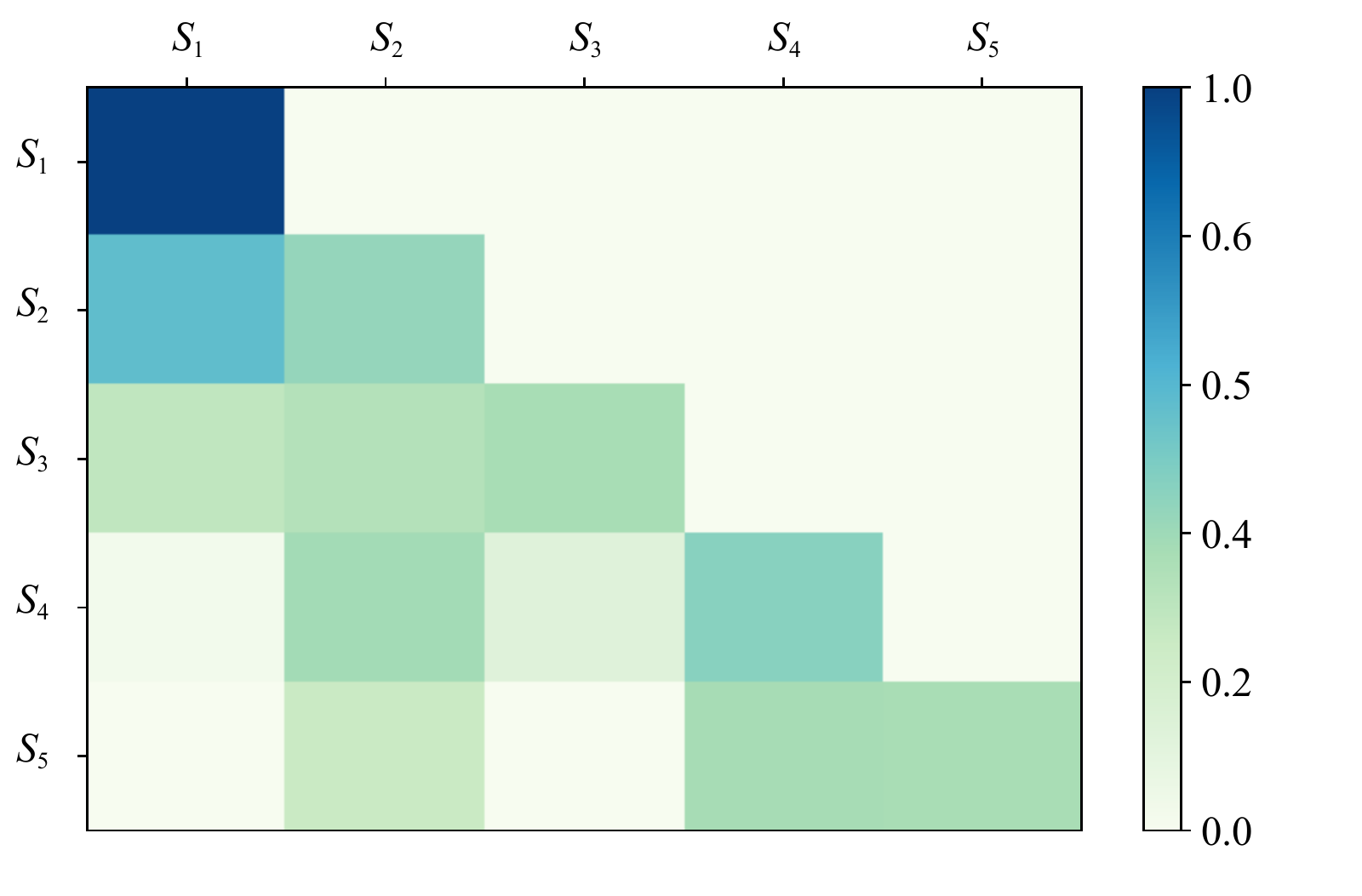}}
\end{minipage}
\begin{minipage}{0.2\textwidth}
\centering
\subfigure[User ID '1492']{\label{b}\includegraphics[width=1.0\textwidth]{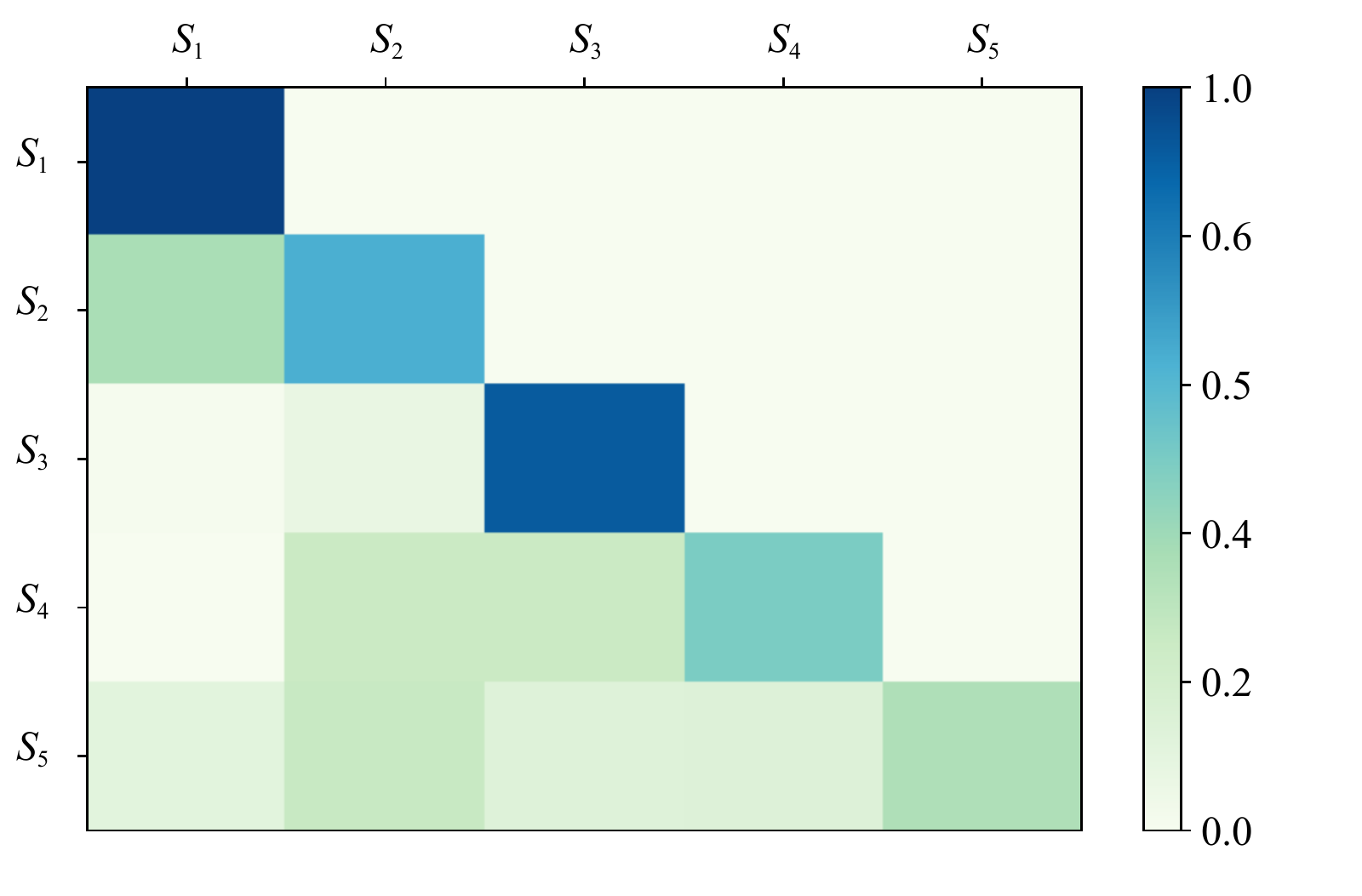}}
\end{minipage}
\caption{Visualization of the self-attention coefficients between sessions.}
\label{fig:inter_weights}
\end{figure}

\begin{figure}[t]
\centering
\begin{minipage}{0.4\textwidth}
\centering
\subfigure[User ID '237']{\includegraphics[width=1.0\textwidth]{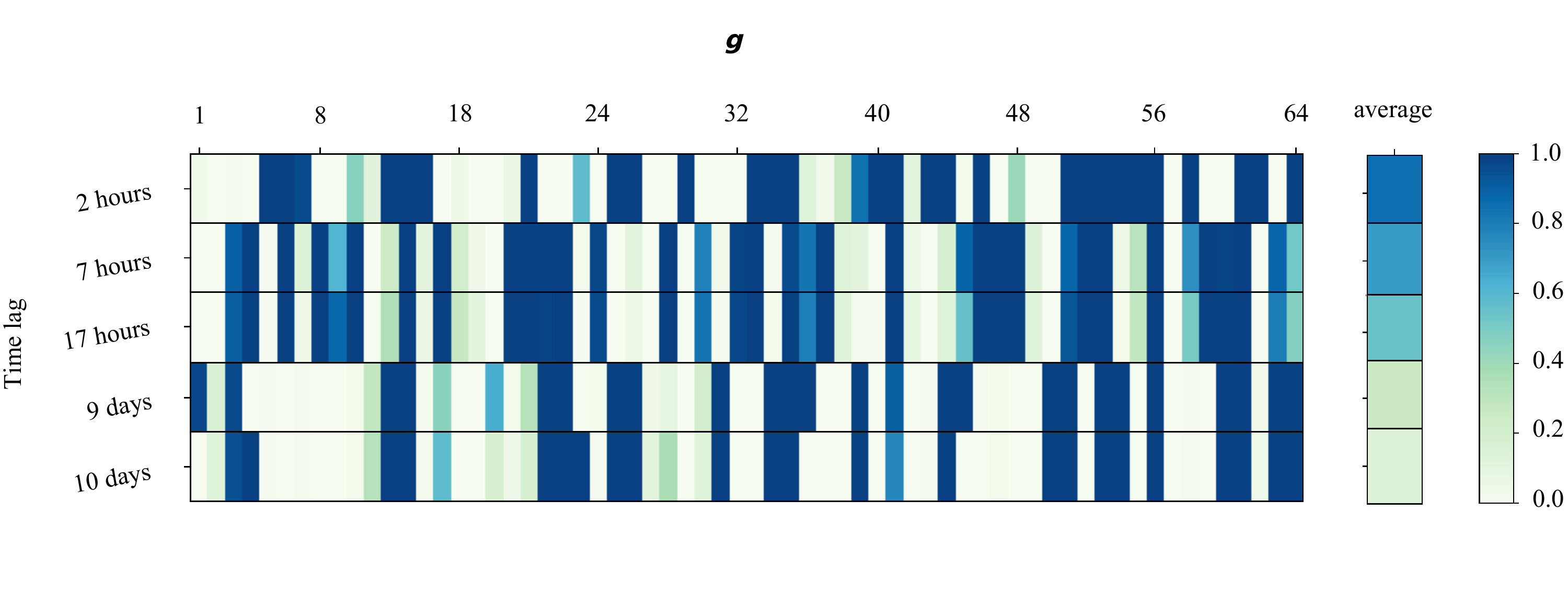}}
\end{minipage}
\begin{minipage}{0.4\textwidth}
\centering
\subfigure[User ID '1492']{\label{b}\includegraphics[width=1.0\textwidth]{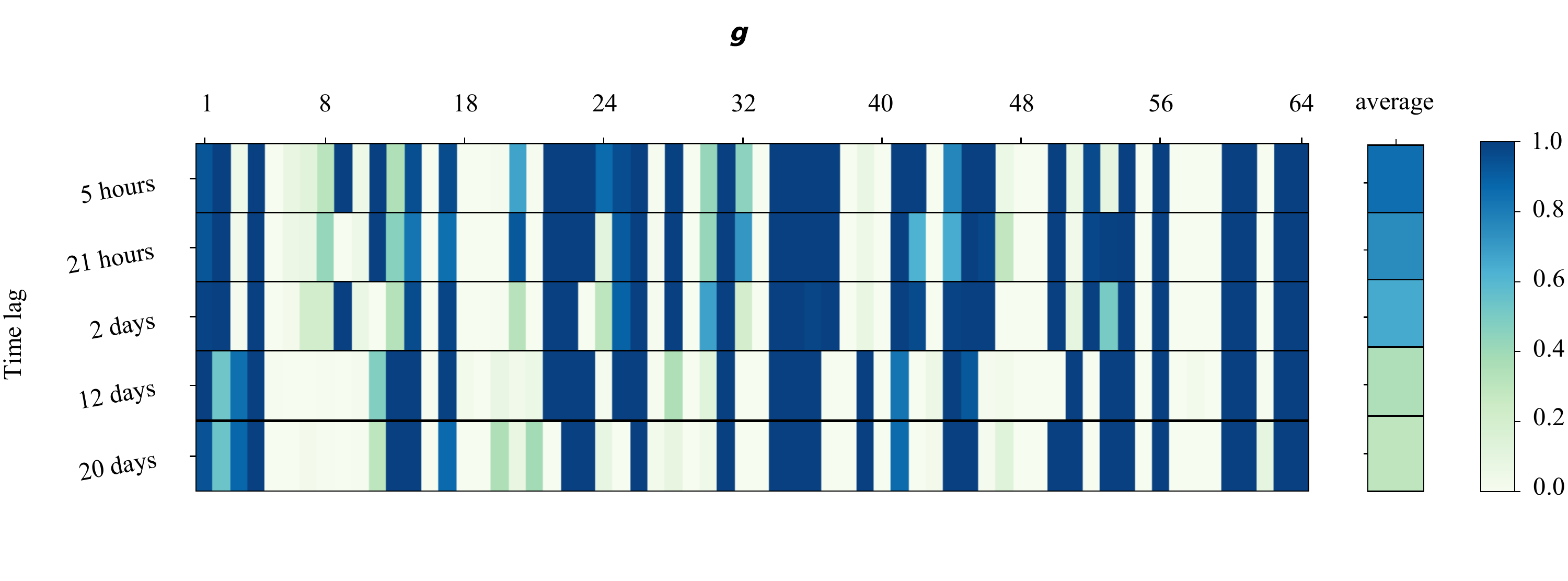}}
\end{minipage}
\caption{Visualization of the time gate vectors.}
\label{fig:gate_vector}
\end{figure}

%\begin{figure*}[t]
%\centering
%\begin{minipage}[t]{0.24\textwidth}
%\centering
%\subfigure[Hit$@10$ over different $T$]{\includegraphics[width=0.8\textwidth]{HIT_Window_Size_T.pdf}}
%\end{minipage}
%\begin{minipage}[t]{0.24\textwidth}
%\centering
%\subfigure[MAP$@10$ over different $T$]{\includegraphics[width=0.8\textwidth]{MAP_Window_Size_T.pdf}}
%\end{minipage}
%\begin{minipage}[t]{0.24\textwidth}
%\centering
%\subfigure[Hit$@10$ over different $d$]{\includegraphics[width=0.8\textwidth]{HIT_d_model.pdf}}
%\end{minipage} \\
%\begin{minipage}[t]{0.24\textwidth}
%\centering
%\subfigure[MAP$@10$ over different $d$]{\includegraphics[width=0.8\textwidth]{MAP_d_model.pdf}}
%\end{minipage}
%\begin{minipage}[t]{0.24\textwidth}
%\centering
%\subfigure[Hit$@10$ over different $h$]{\includegraphics[width=0.8\textwidth]{HIT_h.pdf}}
%\end{minipage}
%\begin{minipage}[t]{0.24\textwidth}
%\centering
%\subfigure[MAP$@10$ over different $h$]{\includegraphics[width=0.8\textwidth]{MAP_h.pdf}}
%\end{minipage}
%\caption{Tuning of hyper-parameters. }
%\label{fig:hyper-parameters}
%\end{figure*}

Now we further illustrate TLSRec's ability to capture the interactions between sessions and its ability to regulate the contributions of long-term preference and short-term preference. For this purpose, we randomly sample two users with IDs '237' and '1492' from Movielens-1M and visualize their self-attention coefficients between sessions and their gate vectors over different time lags in Figures \ref{fig:inter_weights} and \ref{fig:gate_vector}, respectively. 

In Figures \ref{fig:inter_weights}(a) and \ref{fig:inter_weights}(b), one cell $(S_i, S_j)$ at the row $S_i$ and column $S_j$ ($i \le j$) represents the attention given by $S_i$ to $S_j$ that is generated by Equation (\ref{Eq_MultiHeadAtt}), and the darker the color, the greater the attention. From Figure \ref{fig:inter_weights} we can see that there does exist influence between sessions, and to reveal the real preference for a session, TLSRec assigns different attention weights to its previous sessions, by which even the influence of early sessions can be captured.

In Figures \ref{fig:gate_vector}(a) and \ref{fig:gate_vector}(b), a row of the matrices is a time gate vectors corresponding to a specific time lag, along with the average over its dimensions that is shown as the corresponding component in the average column. At first, we can see that the colors of the dimensions of the same time gate vector are different from each other, and again the darker the color, the larger the value. This observation shows that by the time gate vector TLSRec can evaluate the contributions of the short-term preference and long-term preference at the fine-grained dimension granularity for the learning of the current preference, since the $i$th dimension $\boldsymbol{g}(i)$ of the time gate vector and $1-\boldsymbol{g}(i)$ are the weights of the $i$th dimension of the short-term preference embedding and the long-term preference embedding, respectively, during the fusion in Equation (\ref{Eq_Fusion}). From Figures \ref{fig:gate_vector}(a) and \ref{fig:gate_vector}(b), we can also note that the average weight over the dimensions of a time gate vector decays with the increase of the time lag. This result confirms our intuition that the longer the distance between the time of the last behavior and the time when a recommendation is made, the less the impact of the short-term preference of a user on her/his current preference.

\section{Conclusion}

In this paper, we propose a novel model called Time Lag aware Sequential Recommendation (TLSRec). 
To capture the global stability and local fluctuation of a user's preference, TLSRec is able to model a user's long-term preference and short-term preference with a hierarchical self-attention network. Meanwhile, due to the neural time gate, TLSRec can fulfill a fusion of the long-term and short-term preferences with a time lag sensitive regulation at the aspect level for the learning of a user's current preference. At last, the extensive experiments conducted on real datasets demonstrate the effectiveness of TLSRec.

%\newpage

\begin{acks}
This work is supported by National Natural Science Foundation of China under grant 61972270, and NSF under grants III-1763325, III-1909323,  III-2106758, and SaTC-1930941.
\end{acks}

%\balance
\bibliographystyle{ACM-Reference-Format}
\bibliography{TLSRec}


\end{document}